\documentclass[preprint,aps,prb,nobibnotes]{revtex4}
\usepackage{graphicx}
\usepackage{dcolumn}

\newcommand{\bg}{\begin{equation}}
\newcommand{\ed}{\end{equation}}

\begin{document}

\topmargin 0.2 truein
\oddsidemargin +0.1 truein
\textheight 9.0 truein
\textwidth 6.0 truein

\title{
Adaptive Programming of Unconventional Nano-Architectures
}

\author{John W. Lawson}
\email{John.W.Lawson@nasa.gov}
\author{David H. Wolpert} 
\email{David.H.Wolpert@nasa.gov}

\affiliation{
Mail Stop 269-2 \\
Center for Nanotechnology \\
NASA Ames Research Center \\
Moffett Field, CA 94035 \\
}

\begin{abstract}
Novel assembly processes for nanocircuits could present
compelling alternatives to the detailed design and placement
currently used for computers.  The resulting architectures however may
not be programmable by standard means.  In this paper, nanocomputers with
unconventional architectures are programmed using adaptive methods.
The internals of the device are treated as a ``black box" and programming
is achieved by manipulating ``control voltages".  Learning algorithms are
used to set the controls.  As examples, logic gates and simple arithmetic
circuits are implemented.  Additionally, similar methods allow for
recon{fi}guration of the devices, and makes them resistant to certain kinds
of faults.
\end{abstract}

\maketitle

\section{Introduction}
The long-standing trend of miniaturizing electronic components 
is expected to encounter serious obstacles in the not
too distance future. Despite the impressive successes in designing
electronic components down to the atomic and molecular scales, serious
dif{fi}culties are anticipated in assembling them into realistic
functional architectures using current technologies.  One such issue is the
expected, excessive cost of both designing and reliably manufacture chips
on such small scales using conventional methods.
It is possible, in fact, that a major limiting factor in future efforts at
miniaturization won't be related to science or technology issues,
but rather to the high cost of the associated manufacturing processes.

Development of novel approaches for assembling nanochips may be a critical
ingredient for this new technology.  One possibility, for example, could
be to rely on chemical synthesis methods (self-assembly, etc...) to
assemble individual components (which may be individual molecules)
into larger functional unit \cite{tour03}.  Chemical synthesis methods
are capable of organizing large numbers of atoms and molecules into
large, regular, ordered structures.  This might be the basis for a
relatively cheap and ef{fi}cient assembly process. What is not clear,
however, is if any of the resulting structures would correspond to
something resembling a computer chip.  Would we know how to 
program it, for example?

Another potential source of dif{fi}culty is the degree of randomness inherent in
nano-systems.  Traditional computer chips are designed and manufactured
very meticulously, and therefore, are very sensitive to the presence of
defects or disorder. With nanocomputers, the individual components
themselves are expected to be extremely sensitive to disorder, and this
may have a serious impact on their reliability.  Randomness in single
electron transistors, for example, may be unavoidable \cite{lik99}.
The effects of integrating such components into a larger circuit are only
beginning to be considered.

If we are to take advantage of novel assembly
methods and potentially exotic architectures, we will need to learn
how to program the resulting devices.  Some assembly methods may give
very ordered, predictable structures.  Other techniques may be more
extreme and produce a range of essentially random configurations.
In this paper, we consider general purpose methods
to program such devices that are independent
of the internal structure, and therefore also of the assembly process.

Minimally, the devices we consider have input and output leads as well as
a set of additional leads we designate as ``controls".  We assume
that the internal components are sufficiently well-connected to
allow signals to propagate across the device.  Beyond that we consider
the internal structure as unknown.  By
adjusting voltages on the controls leads, we 
manipulate the outputs that appear for a given set of inputs.
In particular, we use adaptive methods to find the set of
control voltages that implement a desired function.  In this way,
we program the device.

To develop and test these methods,
we introduce a simple, abstract model that we call
a randomly assembled computer (RAC).  In this model, we place $N$ two-states
devices (``diodes") randomly on a chip and connect them together with
random strengths.  We designate a subset of the diodes as inputs, outputs,
and controls and attempt to program the RAC by manipulating the controls.
We phrase the programming of such a device as an optimization problem. 
We define an error function that measures the difference between the 
function currently implemented by the RAC and some desired
input-output function $f$.  Since this error function $E_f(\vec{c})$ depends
on the values of the controls, we seek a control vector $\vec{c}$
that minimizes the error.  The task is to find a $\vec{c}$
such that $E_f(\vec{c})=0$.  In general, there may be
many solutions to this equation, or possibly, none at all, for a
particular $f$ and a particular RAC.

It is important to emphasize that our main interest is in developing
programming algorithms.  We are not advocating a particular hardware
architecture or assembly process.  A RAC is therefore only a test bed for our
optimization methods.  In fact, we expect that in the real world, 
there will be a large amount of information available about the internal
structure of the device from the assembly process itself.
For example, in a chemical
assembly process, the device might be imaged.  This information could be 
exploited to facilitate programming/optimization.
Additionally, different degrees of programming can be imagined.  For example,
for very predictable, repeatable assembly processes, adaptive programming
may only be needed once for an entire class of devices.  On the other
hand, for devices with more random internal structures,
individual programming may be necessary.
However, even in this more extreme case, devices could be programmed
in parallel, or possibly be connected together in large number where they
could program each other.  In this work, however,
we do not take explicit advantage of such information.
Thus the problem we set for ourselves may be
more challenging that what will be faced in the real world.  However,
it will provide a good testing ground for our algorithms.

Related work has included ideas using neural nets \cite{nets,bool},
as well as the Nanocell architecture \cite{nanocell}.
The Nanocell approach has some ideas in common with ours,
however, in that approach a particular architecture based
on molecular components is presented,
while we aim to present methods that are independent
of a particular architecture.  Furthermore, many of those results
depended on detailed knowledge and control of the full internal states of the 
Nanocell.  This permitted a ``proof of concept" to demonstrate
that internal states do indeed exist that correspond to certain
simple functions.
The broader issue of how to {\em find} those states was only hinted at, however.
Here, this issue of ``black box programming" is our main concern.

In this paper, we consider different optimization methods to program randomly
assembled computers.  In particular, we will employ simulated annealing
as well as adaptive, multi-agent methods.  We will see that while simulated
annealing is adequate for small systems, larger devices require more 
sophisticated approaches.  We have found in previous work that adaptive
multi-agent methods perform very well for large scale
optimization problems \cite{wotu04}.
Other optimization methods could be considered
as well including genetic algorithms, cellular automata, etc.  These
approaches may have advantages in certain situations, but we found our
methods to be more than adequate for our purposes.
We will be primarily concerned with two general questions.
First, in what generality can RACs be programmed i.e. what is the range of
functionality that can be implemented on a given RAC?
And secondly, what methods might be appropriate to perform the programming?

\section{Randomly Assembled Computers}
\begin{figure}
\begin{center}
\resizebox{110mm}{!}{\includegraphics{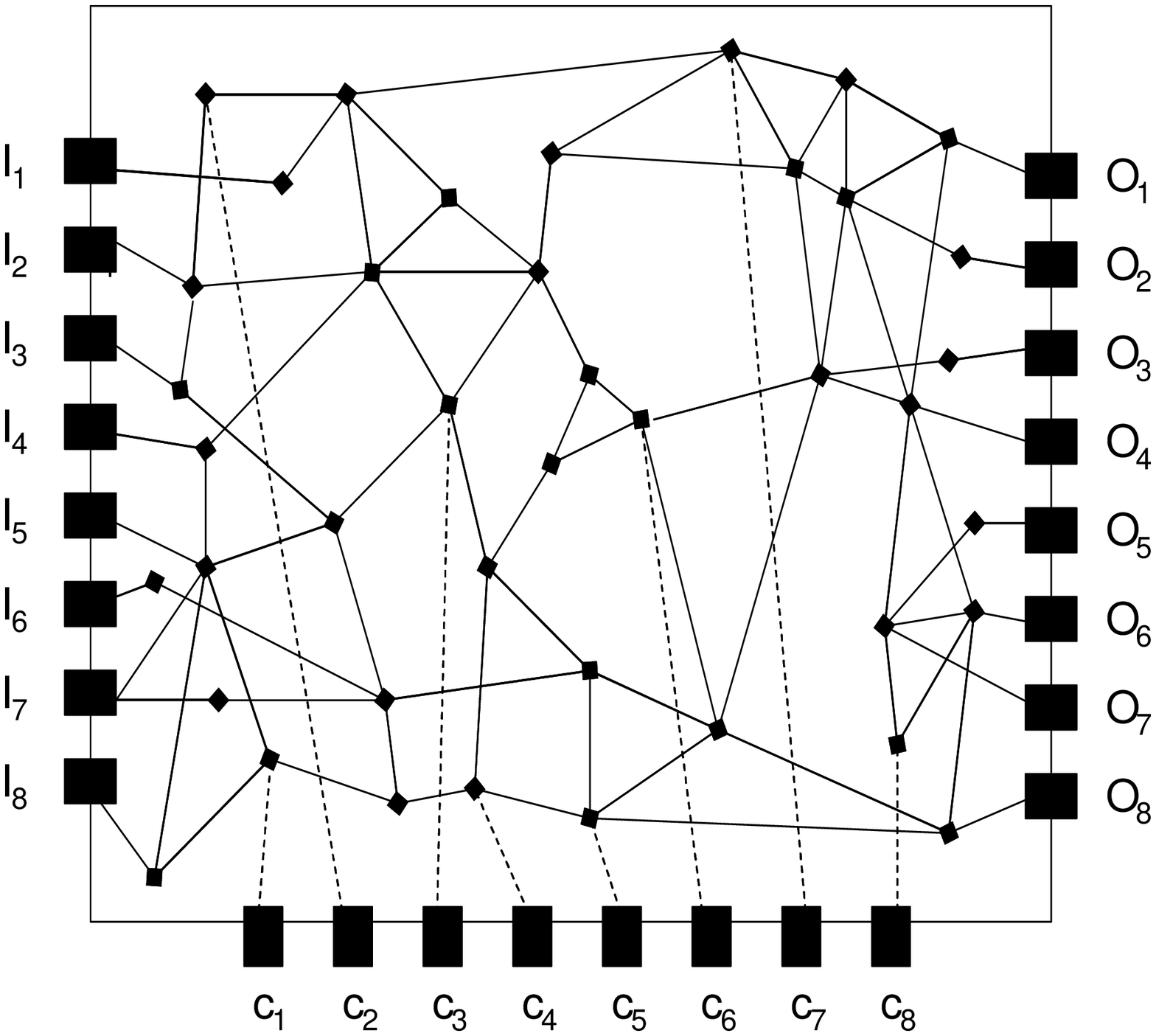}}
\caption{
\label{fig:chip}
Schematic of a Randomly Assembled Computer (RAC).  Diodes 
are placed at random on a chip and connected with random strengths
$\omega_{ij}$.  Connections to the outside world is made via pads
representing inputs, outputs, and controls.  By manipulating control
voltages, different functions can be implemented.
}
\end{center}
\end{figure}

We consider a RAC as having $P$ input variables
$\vec{I}=(I_1,...,I_P)$, $M$ output variables $\vec{O}=(O_1,...,O_M)$, and $K$
control variables $\vec{c}=(c_1,...,c_K)$.  A schematic of a RAC can be seen
in Figure~\ref{fig:chip}.  Here input and output variables
are binary-valued, while the controls can take continuous values.  Our goal
is to {fi}nd a $\vec{c}$ that implements a desire mapping
\bg
f:\vec{I} \rightarrow \vec{O}.
\ed
We specify a ``target'' function $f$ as the ordered set of outputs associated
with all possible input vectors, and where $\vec{T}=(T_1,...,T_M)$ are the
target outputs.

A RAC is properly ``programmed" when $\vec{c}$
causes the RAC outputs $\vec{O}$ to equal the targets $\vec{T}$ over a
``training set" of $L$ input/output pairs $\{\vec{I}^l,\vec{T}^l\}$
representing $f$. We quantify how well we have programmed a RAC thru an
error function de{fi}ned as
\bg
E\equiv E_f(\vec{c})=\frac{1}{ML} \sum_{l=1}^{L} || \vec{O}^l
- \vec{T}^l ||
\ed where $\vec{O}^l$ is shorthand for $\vec{O}_{\vec{c}}(\vec{I}^l)$
and $|| \vec{O}^l - \vec{T}^l ||$ is the number of components
in which $\vec{O}^l_{\vec{c}} (\vec{I}^l)$ and $\vec{T}^l$ disagree.
Thus, $E$ is the fraction over all output variables and training set elements
of mistakes made by the RAC.
Programming a RAC corresponds to minimizing $E$.
In general, the results we will present will be averages over large
ensembles of RACs.

To study different programming methods, the following model
system was used \cite{womil03}.  The internal structure of a RAC
was represented by randomly
placing $N$ ``diodes" 
$d_i(t), i=1,...,N$ (i.e., two-state devices such that $d_i\in [0,1]$)
on a chip, and
connecting them together with random strengths.  An arbitrary 
subset of diodes was designated as inputs, and another subset (perhaps overlapping) as
outputs.  Diodes can change state based on their individual inputs at discrete
time steps $t$ governed by a global clock.  Thus, a RAC has the structure of
an iterated function $\vec{d}(t+1)=F[\vec{d}(t)]$.
A computation begins at $t=0$ by setting each
input diode $d_{p'}(0)=I_p$ where $I_p$ is the input to the RAC
and $p'$ is the diode associated with input bit $p$.
All other diodes are initialized to zero. At subsequent
time steps, the diodes update their states according to the rule
\bg
d_i(t+1)=g(\sum_j^N \omega_{ij} d_j(t) + \sum_j^K \rho_{ij} c_j(t))
\ed
where the $\omega_{ij},\rho_{ij}$ are pre-{fi}xed random numbers, and
$g(x)=\frac{1}{2}(1+tanh(\beta x))$ is a smoothed step function with
``gain" $\beta$ that models the switching behavior of the diodes.
Here, the $\omega_{ij}$ represent the strength of the connection between
two diodes, $i$ and $j$, analogous to a resistance.  Note also that for low values of the gain, 
the diodes can take on values between zero and one, given by the function $g(x)$.
After a {fi}xed number of time steps $t=T$, the outputs
can be read to give the result of the computation $O_m=d_{m'}(T)$,
where $m'$ is the diode associated with output bit $m$.
For simplicity, we consider only circuits composed of diodes and do
not include other possible components such as memory elements, etc.

It should be noted that our model is used to test our programming approach.
However, none of the adaptive programming methods that we present here depend
on the particular details of the model, and are expected to be general enough
to work for any ``black box" nano-circuit.  Moreover, our model captures
several important features of real circuits consisting of an iterated network
of diodes whoses states can change depending on their local environment.
It is also important to note that while the $d_i(t)$ evolve according to a
dynamical rule, Equation 3, the $c_i(t)$ are external, and potentially
time-dependent parameters.  Our task, in fact, will be to set these parameters.

\section{Multi-Agent Methods}
We want to {fi}nd the best $\vec{c}$ that minimizes $E_f(\vec{c})$
for a given training set.  The best case scenario is to {fi}nd a $\vec{c}$
that gives $E_f(\vec{c})=0$, i.e., {fi}nd a $\vec{c}$ so that the RAC gives
the correct outputs for the given inputs over all the examples.  We will
see that this is often achievable.  Several issues make this a potentially
dif{fi}cult optimization problem.  First, $E$ is a very nonlinear function
with a potentially large number of local minima. Secondly, the exact
functional form of $E_f(\vec{c})$ is not known due to our desire to treat the
RAC as a black box. The only information that we have available is the set
of input/outputs pair values $(\vec{I}^l,\vec{O}^l)$ for a given control
$\vec{c}$. Lastly, there may be a large number of controls that need to be set.
Because of these dif{fi}culties, we will employ a set of methods recently
developed in the context of multi-agent systems.  These multi-agent methods
have some relation to simulated annealing (SA), but have signi{fi}cant
differences, and have shown dramatically faster convergence in a number
of problems.

There are three main differences between the multi-agent methods we will
consider and more conventional methods such as Simulated Annealing (SA)
\cite{wotu04,wowh00,wotu99c,wotu01a}.
First, the multi-agent approach is distributed.  Thus, instead of considering
a $N$-dimensional optimization problem, we deal with $N$ 1-dimensional ones.
Each independent variable $c_i$ is regarded as an agent, and each agent $i$
separately sets its variable $c_i$ in order to try to optimize an associated
objective function $e_i(f, \vec{c})$.
Thus, we have $N$ independent optimization processes, each of which we are
calling an ``agent".
In general, distributed approaches
such as these are expected to scale better for large problems.

Secondly, each agent solves its particular optimization problem using
Reinforcement Learning (RL) \cite{suba98}.  The version of RL we use is
called Boltzmann learning and is related to SA.  The principal difference is
that instead of taking ``random" trial steps, RL relies on previous data to
make ``smart" trial steps.  In this way, RL algorithms converge very rapidly.
Boltzmann learning is also easy to implement.  More sophisticated algorithms
may be necessary, however, for more dif{fi}cult problems.

In our RL algorithm, at each simulation time step $\tau$, each agent $i$
randomly generates a number of candidate values for $c_i(\tau+1)$ from some
$\Delta$ neighborhood of the current $c_i(\tau)$.  Next, the agent estimates,
based on data from previous time steps, a value of $e_i$ for each candidate
value.  This is done by performing a weighted average over all previous pairs
$(c_i(\tau'),e_i(\tau'))$, for $\tau'<\tau$.  The weighting damps the
contribution from ``old" data (i.e., from $c_i(\tau')$ and $e_i(\tau')$
where $\tau'<<\tau$).  Finally, a Boltzmann probability distribution over
these estimates is sampled to select the resulting trial move.  A nonzero
simulation temperature prevents us from getting stuck in local minima.

After the agents have all made their moves, it may turn out that the
global error $E$ has actually increased.  This may happen if, for example,
the agents' estimates are inaccurate.  To address this, we also
include a Metropolis-style global accept/reject step.  If after the
agents have made their moves $E$ has decreased, then we accept the
new $\vec{c}$. Otherwise we reject it with conditional probability
proportional to $~e^{-\beta (E-E')}$ \cite{wotu04}. 

The third difference with conventional methods is that the individual agents may be
assigned different objective functions.  Since our global objective is to
minimize $E$, we might expect that each agent should use $E$ as its individual
objective function, $e_i=E$.  Such a system of agents who all have the same objective
function is called a Team Game (TG), in analogy with the scenario in game
theory where all the players/agents have the same payoff function.
We expect --- and indeed observe --- that for small systems Team Games
outperform SA. This is because by using learning algorithms, the agents can
make ``smart" trial moves inferred from their past history, as opposed to the
random ones with conventional SA.

However, as the system size grows, it becomes more dif{fi}cult for each
individual agent to discern its impact on $E$, and thus it becomes more
dif{fi}cult for each agent to choose an optimal value if $e_i = E$. To
address this we de{fi}ne, heuristically, a signal-to-noise measure for each
agent $i$ that we call ``learnability": 
\bg
\lambda_i^{e_i} = \frac{|\partial_i e_i|}{|\vec{\nabla}_{\hat{} i} e_i|}
\ed
where
$\partial_i e_i = \partial e_i/\partial c_i$, and $\vec{\nabla}_{\hat{} i} e_i$
denotes the gradient of $e_i$, but with the component in the $i^{th}$
direction removed. The {\em learnability} of agent $i$ measures
the {\em sensitivity}
of $e_i$ to changes in $c_i$ relative to changes in the system as a whole.  
It is expected that for large $\lambda_i$, agent $i$ can more easily discern
its impact on $e_i$, and therefore, it can make better trial moves. On the
other hand, we expect that beyond a certain system size, the denominator
of $\lambda_i^{e_i}$ will grow to such a point that the agents will be
essentially making random moves. At that point, Team Games and SA will
basically give the same result.

Even though our goal is for the system as a whole to minimize the error $E$,
there is nothing to prevent us from giving different $e_i$ to the different
agents. To exploit this, each agent is assigned a ``difference utility" $e_i$
of the form
\bg
e_i(\vec{c}) = E(\vec{c}) - E(\vec{c})|_{c_i=0}.
\ed
Since $\partial_i e_i = \partial_i E$, critical points of $e_i$ will be
critical points of $E$.  Thus, if agent $c_i$ optimizes $e_i$, it will
optimize $E$ as well. The functions $e_i$ that are {\em aligned} with $E$ in this
manner are called {\em factored}.  

In addition to being factored, $e_i$ also typically has better learnability
than $E$.  This is because the numerators of $\lambda_i^E$ and $\lambda_i^e$
are identical, but the denominators are very different.  In fact, we expect
in general that $|\partial_j e_i| << |\partial_j E|$.  This should lead to a
substantial reduction in the background noise, and hence an increase in
$\lambda_i$. This behavior is born out by many simulations.  Such $e_i$
that are both ``factored" and have high ``learnability" are expected to
perform well for large systems.

Notice that equation (5) is only one choice for a difference utility.
In previous work, this utility was called the Wonderful Life Utility
(WLU). The value for $c_i$ in the subtracted term could have been set
to a different, non-zero, value, and the advantages would still
hold. A similar possibility is to subtract an expectation value taken
over all $c_i$ weighted by an appropriate probability density.  \bg
e_i(\vec{c}) = E(\vec{c}) - < E(\vec{c}) >_{c_i}.  \ed Previously,
this utility has been called the Aristocrat Utility (AU). Both WLU,
AU, and other utilities have been studied extensively \cite{wotu01a}.
In this paper, we will concentrate on WLU for simplicity. 
A fully formal derivation of WLU and AU based on bias and variance may
be found in \cite{woinfo,wobien}.  This work has many other advantages
beyond that presented here, e.g., explicit connection to bounded
rational game theory and statistical physics. However it is more
involved than is needed for current purposes.

\section{Results}
We performed simulations on a variety of target functions $f$, and on RACs of different
sizes (N,K), where N is the number of diodes and K is the number of controls.
Once a RAC was assembled and a function selected to be programmed, an error
function could be de{fi}ned.  Depending on the size of the RAC, we use different
optimization algorithms to program them. For smaller RACs $(N,K < 20)$, SA
often proved adequate, but for larger RACs $(N,K > 50)$, multi-agent methods
were essential.

To begin, we {fi}rst considered 2-input, 1-output logic gates: INV, AND, OR, and XOR
programmed using simulated annealing for the optimization.  
For these simple cases, SA performs adequately, and in general, runs
faster (in minutes as opposed to hours on a workstation)
than the multi-agent approaches.
We demonstrate that not only can these functions indeed be
programmed, but we attempt to answer an important general
question which is what is the range of functionality that
a given RAC can implement.
Indeed, we expect that some functions cannot be implemented at all on a
particular RAC.  We might hope, however, that a variety of functions can be
programmed on a RAC of suf{fi}cient size.  That minimal size will depend on
the particular function being considered. Thus, even for logic gates, we might
ask:  how big of a RAC is needed to have a high programming success rate?

\begin{figure}[tbp]
\resizebox{120mm}{!}{\includegraphics{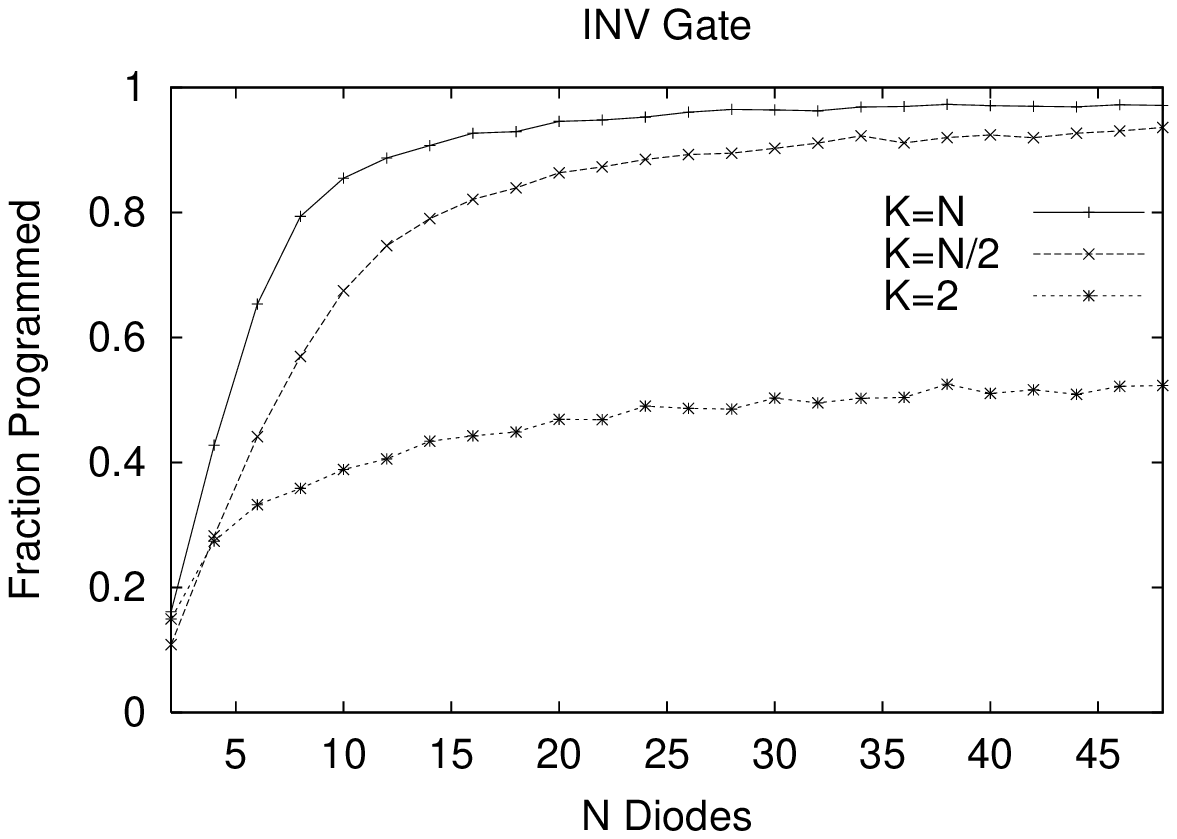}}
\resizebox{120mm}{!}{\includegraphics{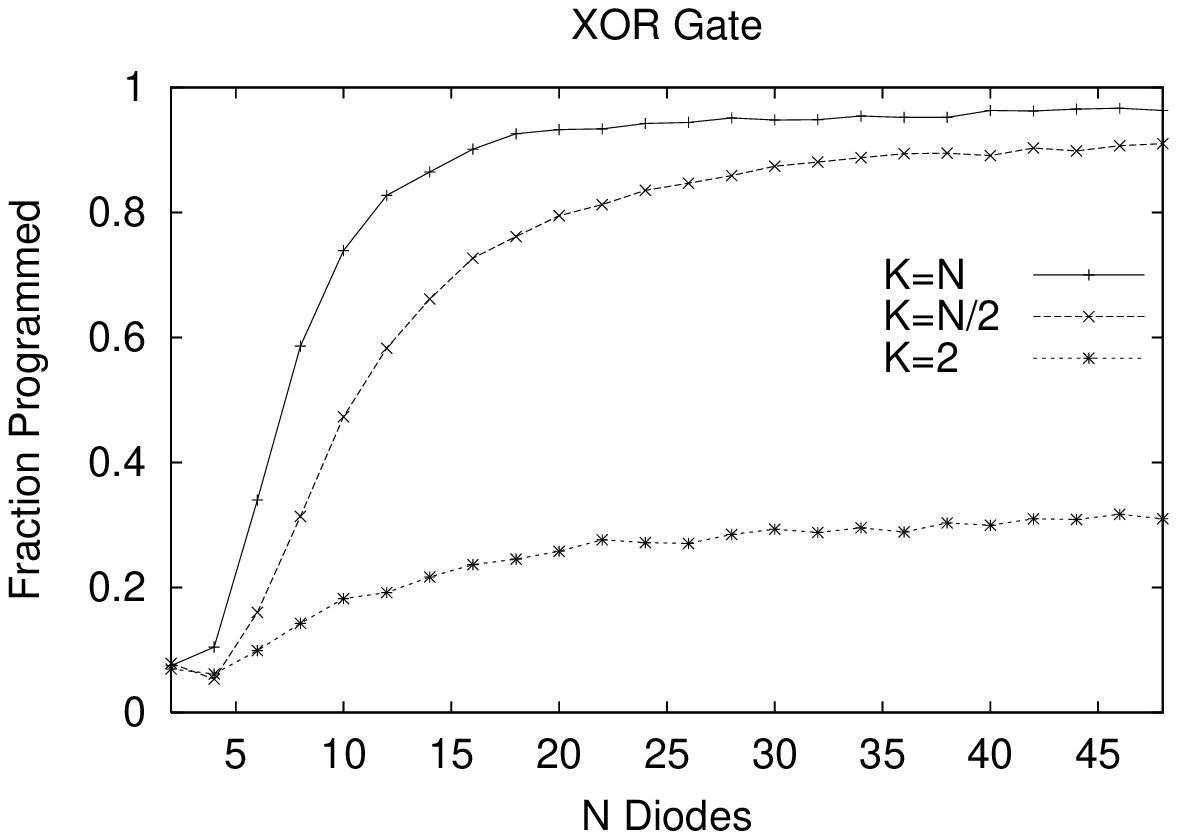}}
\caption{
\label{fig:inv}
Fraction of RACs with $N$ diodes and $K$ controls 
successfully programmed to implement INV and XOR function.
The fraction is out of an ensemble of 10,000 RACs.
Simulated annealing was used for the programming.
}
\end{figure}

The results for certain logic gates are shown in Figure~\ref{fig:inv}.
Each data point represents the result of programming attempts on 10,000
different RACs of the given size $(N,K)$.  The y-axis gives the fraction
that were successfully programmed i.e. all the outputs matched the targets
with no mistakes ($E=0$).  Different curves correspond to different proportion 
of controls to diodes.  We see that the Inverter (INV) with $K=N$ has a greater
than $90\%$ success rate for $N\geq12$ whereas the XOR required $N\geq18$.
This is expected due to the greater complexity of the XOR.  Furthermore,
we see that reducing the number of controls by half has only a relatively
small effect on the performance, with the success rate still better than
$90\%$, but for $N\geq30$.  Lastly, we see, even with a fixed number of 
controls (K=2), a large fraction ($50\%$ for the INV and $30\%$ for the XOR)
can still be programmed.

Next, we considered larger circuits and larger RACs. We implemented larger
functional units, namely arithmetic circuits.  In particular, we examined
2-bit adders with carry and multipliers.  These are signi{fi}cantly more
complicated functions than the 2-input logic gates.  Our adders and multipliers
performed operations on two 2-bit numbers, giving a total of 4 input bits,
and 3 output bits.  With 4 inputs, there are 16 possible input combinations,
and in total, $2^{3 \times 2^4}=2^{48}$
possible binary functions from 4 inputs to 3
outputs.  An adder or a multiplier is only one of these possible functions.
Our optimization algorithm was required to {fi}nd these particular functions
out of the large number of possibilities. For these functions, with 16 possible
inputs and 3 bit output, there are 48 total output bits that must be correctly
set in order for the RAC to have been programmed perfectly.  The
truth table in Table 1 gives the required input/output combinations that de{fi}ne
these functions.

For functions of the complexity of adders and multipliers, we used RACs of
size (N=100,K=40) with (T=5) iteration cycles. For a dynamically controlled
RAC (i.e. the control voltages change with each clock cycle
$\vec{c}(t)$), this results in 200 control parameters.  For such a
large number of parameters, multi-agent methods were essential. 

\begin{figure}[tbp]
\resizebox{120mm}{!}{\includegraphics{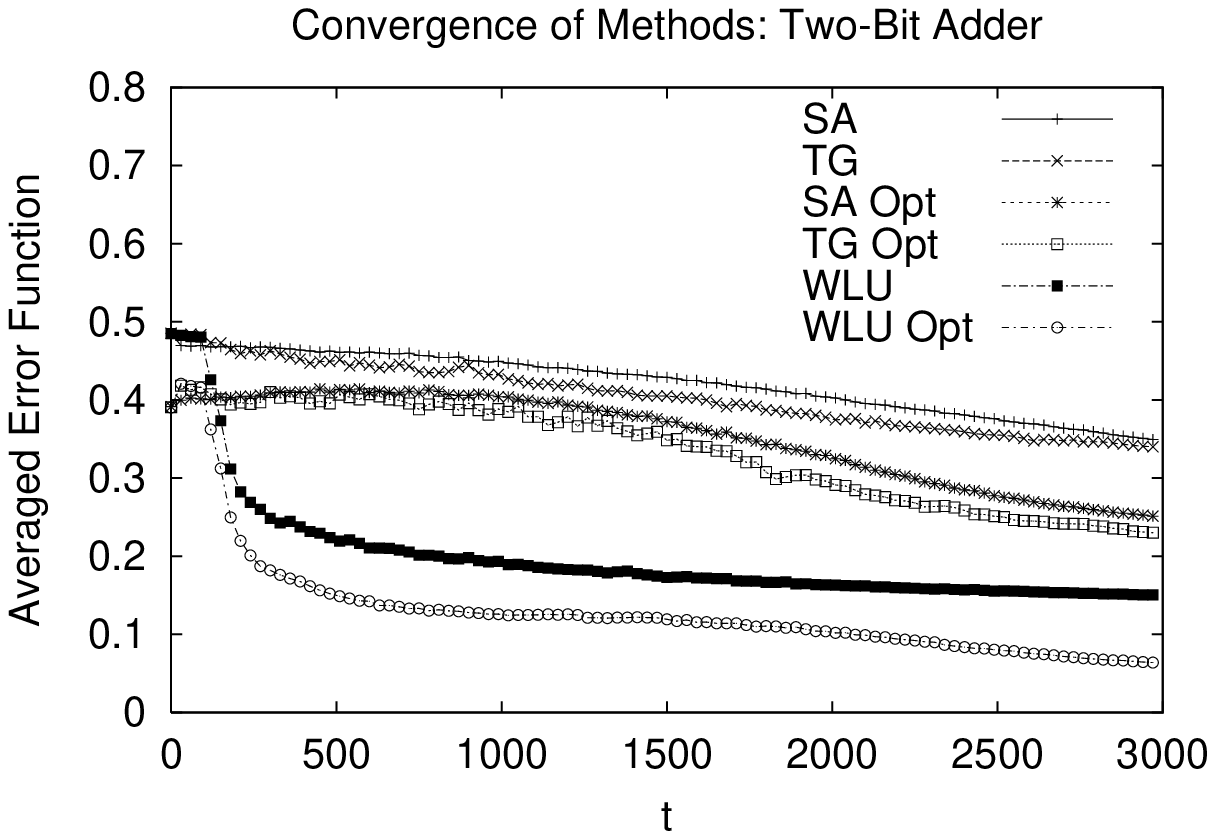}}
\resizebox{120mm}{!}{\includegraphics{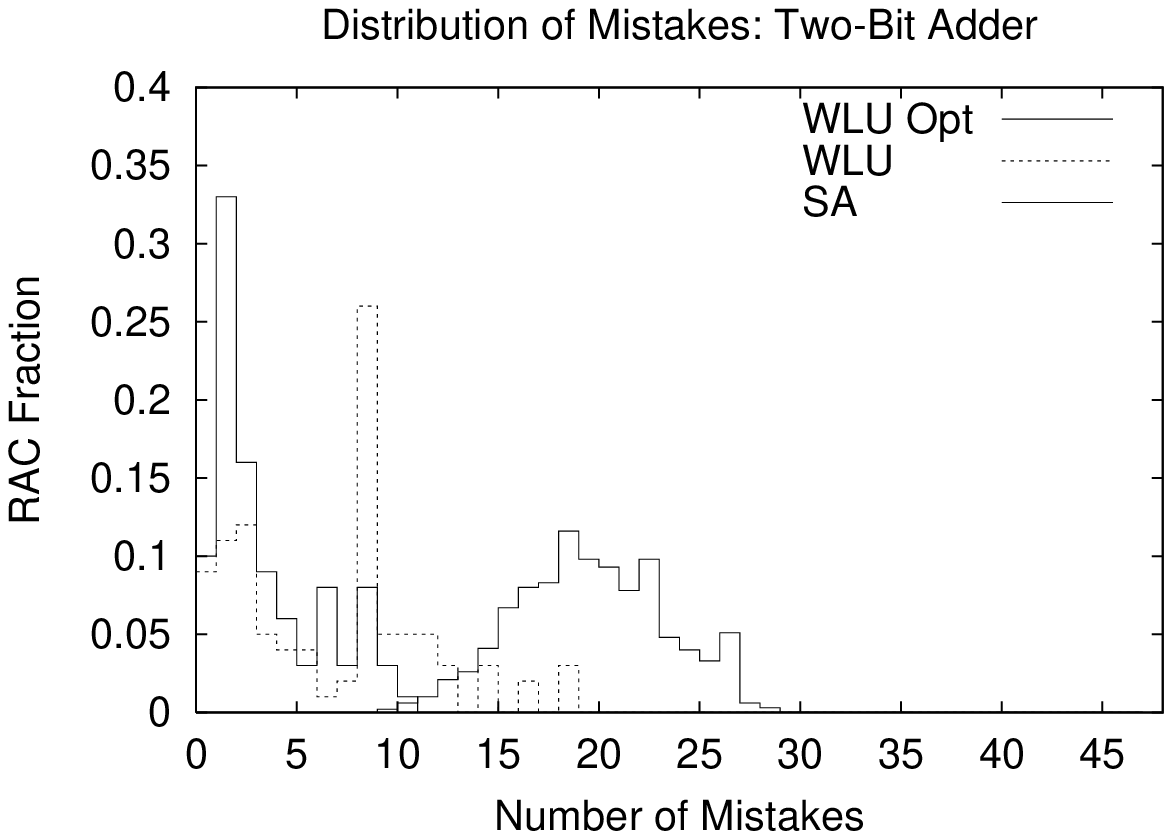}}
\caption{
\label{fig:add}
Two-Bit Adders with carry with
N=100 diodes, K=40 controls, and T=5 iterations.
Top plot compares the convergence of the error function averaged
over 1000 RACs using different programming methods.
Bottom plot gives the fraction of RACs making a
given number of mistakes (out of 48 possible) {\em after} programming.
WLU Opt performs the best with $10\%$ making no mistakes
and $33\%$ making one mistake.  WLU is next with $9\%$,$11\%$, and $12\%$
making 0,1, and 3 mistakes.
The best result for SA is $0.2\%$ making 9 mistakes.
}
\end{figure}

\begin{figure}[tbp]
\resizebox{120mm}{!}{\includegraphics{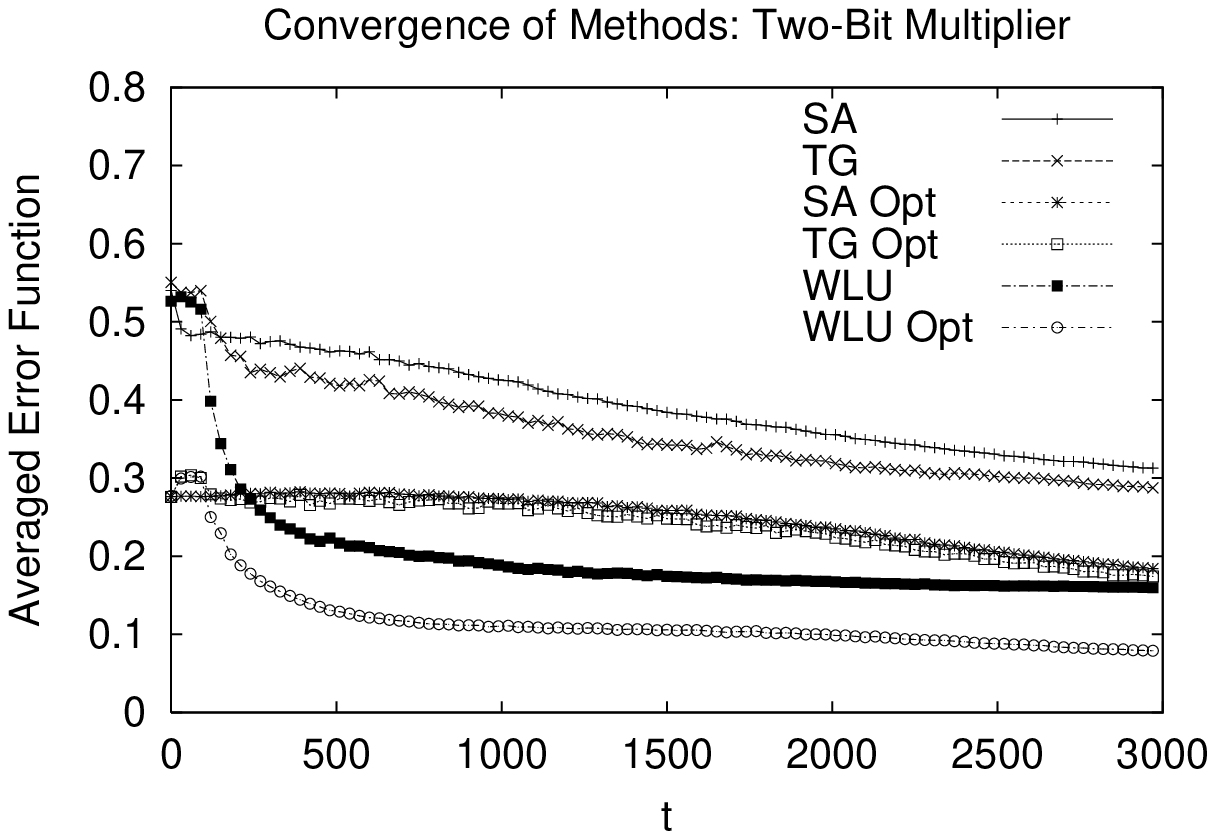}}
\resizebox{120mm}{!}{\includegraphics{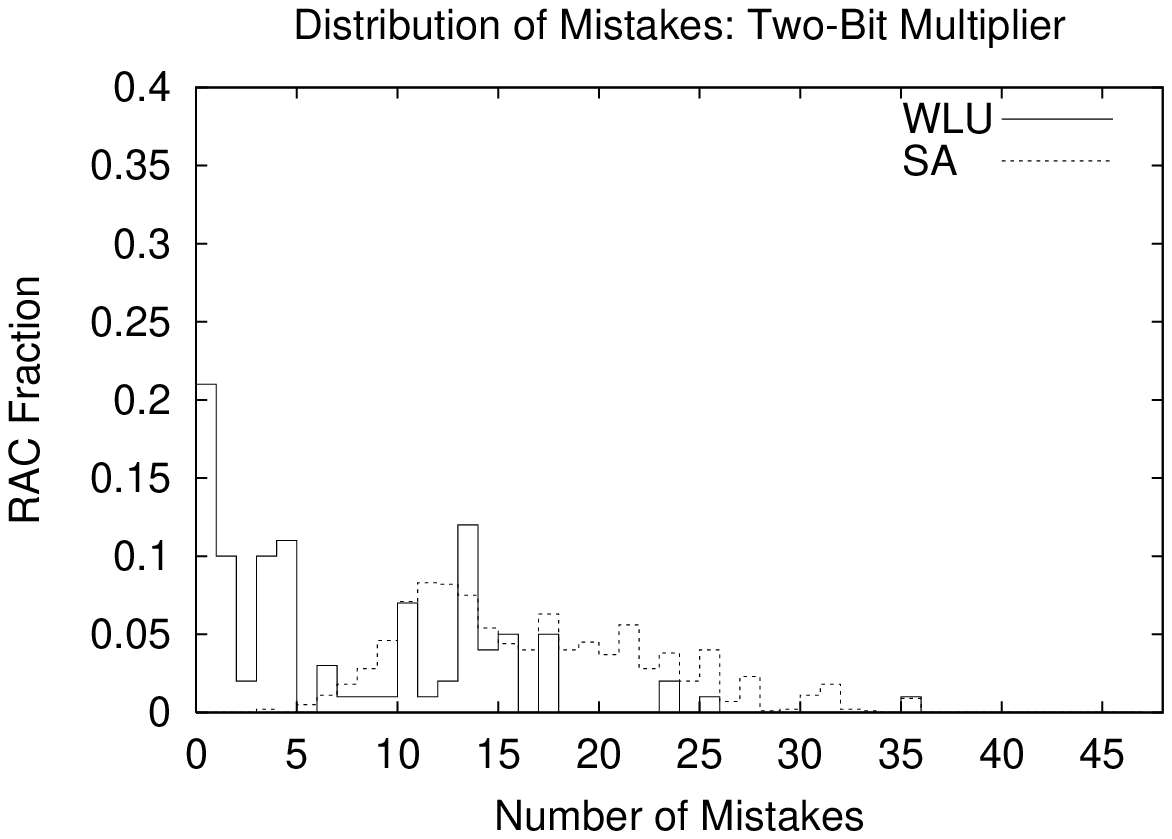}}
\caption{
\label{fig:mult}
Two-Bit Multiplier with N=100 diodes, K=40 controls,
and T=5 iterations.  WLU programs $21\%$ with no mistakes and SA's best
effort is 3 mistakes.  In this case, WLU Opt has a smaller fraction 
than WLU making
zero mistakes at $9\%$ even though its ensemble averaged error function
is lower.  For clarity, only WLU and SA are shown.
}
\end{figure}

The upper plots in the Figures~\ref{fig:add},~\ref{fig:mult} for the
adder and the multiplier show the
convergence of different methods as a function of simulation time.
In each case, the error function $E_f$ is averaged over an ensemble of 1000 RACs.
We see for both the adder and the multiplier, that for a RAC of this size, the Team Game (TG)
behavior is only marginally better than SA, while the multi-agent WLU does considerably better
due to its superior scaling properties.
In the lower plots,
we deconstructed the convergence graphs and show how well the
individual RACs were programmed.
Each data point gives
the fraction of RACs that made a given number of mistakes after programming
was completed.  The ideal situation would be a fraction of 1 (all the RACs)
making 0 mistakes;
this would correspond to all the RACs being perfectly programmed. We see that
WLU programmed RACs made much fewer mistakes than SA programmed ones.
In fact, for the adder,
almost $50\%$ of the RACs programmed with WLU made only 0 or 1 mistake out of
48.  In no case, were there more 10 mistakes with WLU.
On the other hand, the best performing SA RAC still made 9 mistakes.
For the multiplier, more that $20\%$ of the WLU RACs make no mistakes while the
best for SA is at 4 mistakes.

For these larger circuits, we did additional
runs where we show how including additional 
information or control over the circuit can affect the
performance.  In programming unconventional
circuits, some information may be available about the internal structure.
For example, with a random assembly process,
large batches of chips might be produced very inexpensively due to the 
negligible design and placement costs.
It would be very easy therefore to quickly generate and
test a subset of these devices
to see which functions they implement without any programming all,
and then pick the ones whose initial error was the lowest
for subsequent programming.
Clearly, it would more ef{fi}cient
to program a RAC that is closer originally to the desired function.
To model this possibility, we generated 100 RACs, and programmed
the one with the lowest initial error.
Additionally, it may also be possible that during the programming process,
we may have control over some physical parameters of the device
such as an external field.
To model this scenario, we adjust the gain parameter $\beta$ of our model
during the programming process.
In particular, we found an improvement in performance by annealing the gain.
Sets of runs with these additional features are labeled ``Opt" in
Figures~\ref{fig:add},~\ref{fig:mult}.  In all cases, the ``Opt" runs
improved performance.

\begin{figure}[tbp]
\resizebox{90mm}{!}{\includegraphics{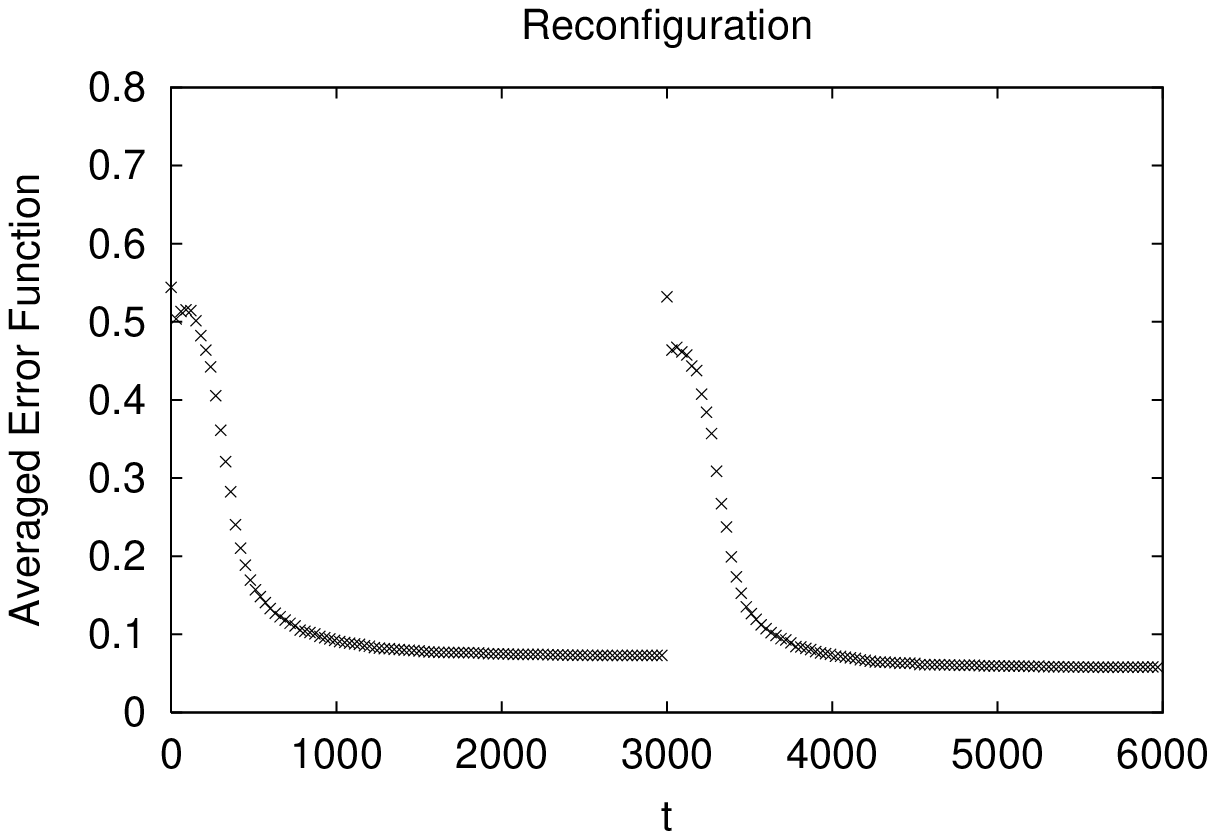}}
\resizebox{90mm}{!}{\includegraphics{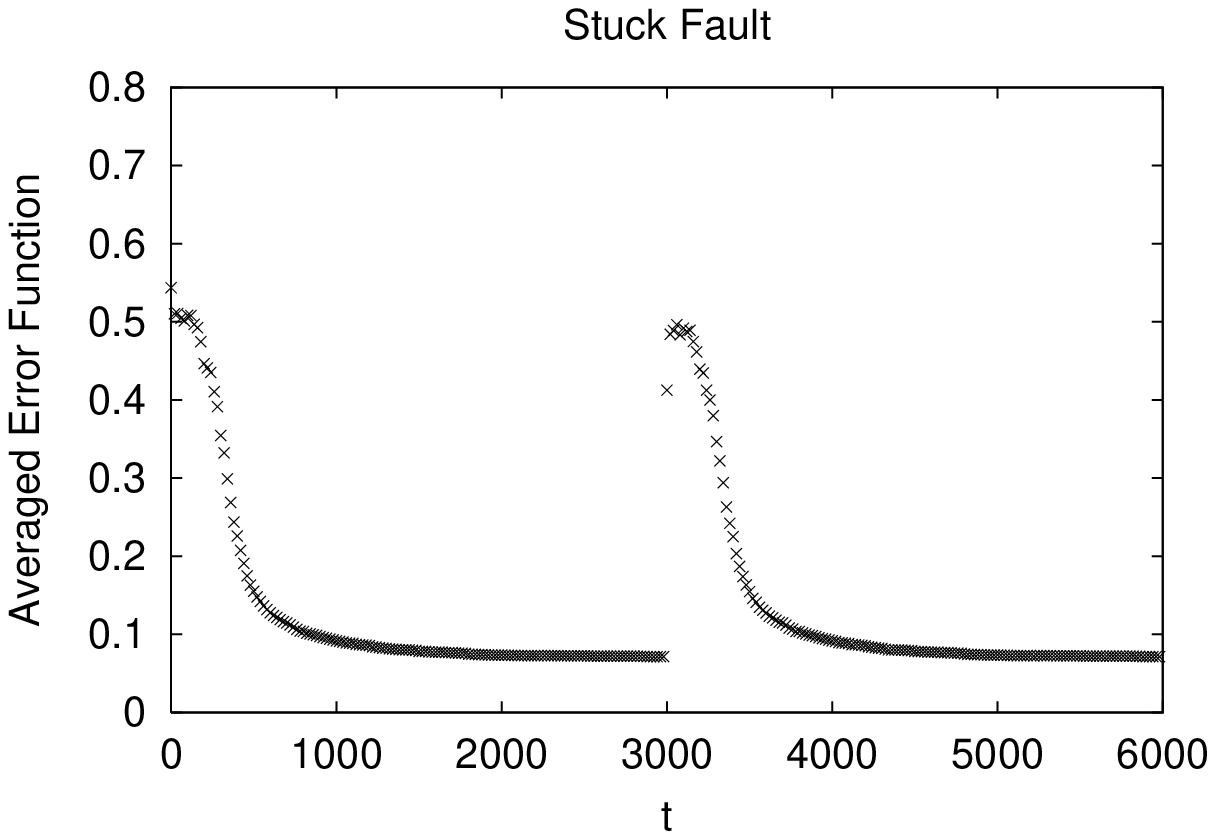}}
\resizebox{90mm}{!}{\includegraphics{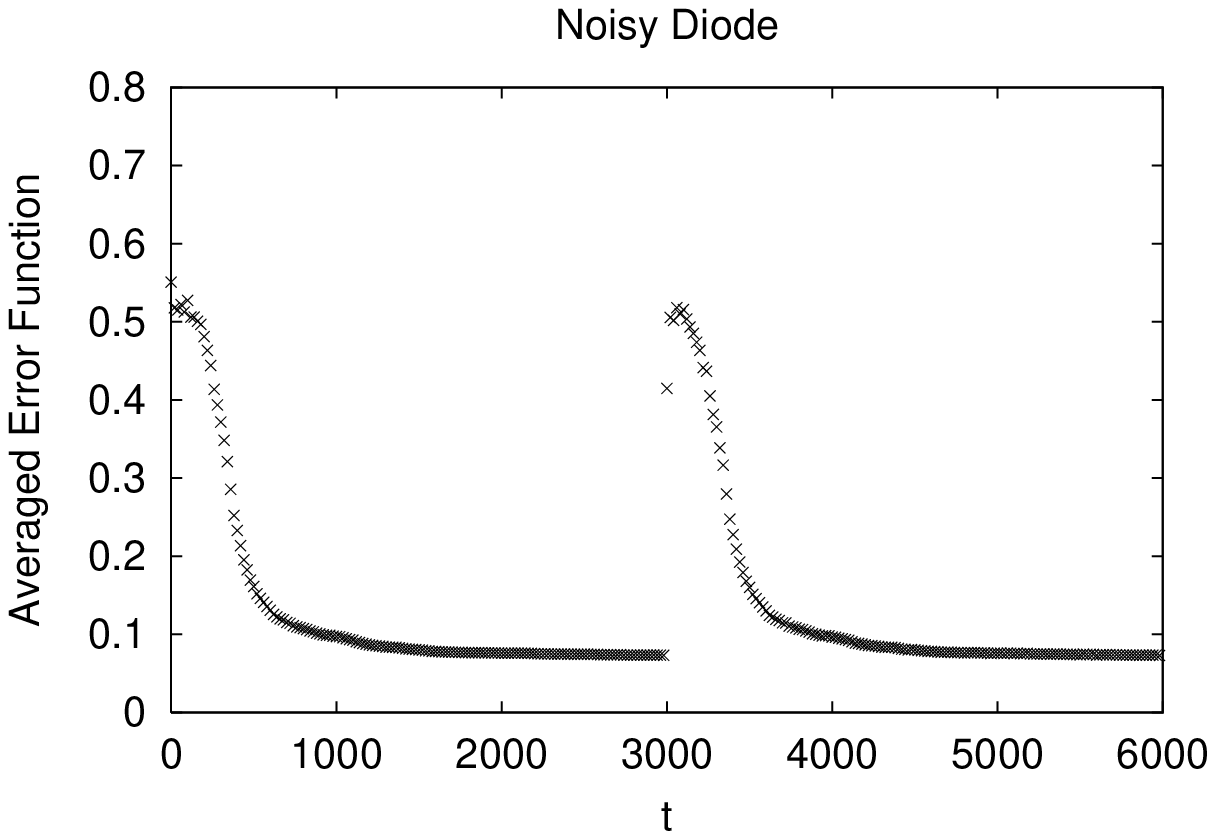}}
\caption{
\label{fig:apps}
Adaptive approach to fault tolerance and reconfiguration.
Error functions is averaged over 1000 RACs with (N=20,K=10).
An AND function is initially implemented.
At t=3000, the system is perturbed.
In the top plot, the RACs are recon{fi}gured to implement 
an XOR function.
In the middle plot, a ``stuck" fault is introduced, and in the bottom plot,
the RACs develop a ``noisy" component.
In each case, the RACs adjusts to the new situation.
}
\end{figure}

The random assembly of the RAC and the adaptive nature of the multi-agent
methods makes them well-suited for handling faults, defects, noisy
components, and also allows them to be redesigned/reused.  To illustrate this
{fl}exibility, we performed the following simulations.  First, to illustrate RAC
reuse/recon{fi}gurability, we initially programmed an ensemble of RACs
with $(N=20,K=10)$ to implement AND functions.  From Figure~ref{fig:inv}, we 
expect a $90\%$ success rate across the ensemble.
Then, at simulation time step $\tau=3000$, we
abruptly changed the same RAC to implement a different function, namely an XOR.
As can be seen in the top plot of Figure ~\ref{fig:apps}
which is averaged over 1000 RACs, the RACs are able
to adjust quickly to the new functionality.

Secondly, we considered recovery from a fault.  Initially, we programmed the
RACs to implement ANDs.  Then at $\tau=3000$, we randomly chose a diode and
{fi}xed its value at $d_i=1$ for the remainder of the simulation.  This was to
simulate a ``stuck" fault.  We see that initially there is a large increase 
in the error due to the fault, but then the adaptive algorithm was able to
program around it.  Note that the $\vec{c}$ that implements the AND, before
and after the fault will, in general, be very different.

Finally, we
considered the case of a ``noisy diode".  To simulate this, we again programmed
a group of RACs to implement AND functions.  At $\tau=3000$, we randomly chose
a diode to become noisy.  At each simulation time step, its value was {fl}ipped
at random.  This is in contrast to the stuck fault were the diode maintained
a {fi}xed value.  We see that even though this diode is continually changing
state, the adaptive algorithms are able to effectively neutralize it,
and again implement a AND function.  These results are shown in
Figure~\ref{fig:apps}.

\section{Summary}
In this paper, we considered programming nanocomputers with unconventional,
and potentially unknown and/or random, architectures.  Novel assembly
processes may represents potentially inexpensive alternatives to the high
costs anticipated to extend conventional design and manufacturing methods
to the nanoscale.  The resulting architectures may not respond to 
conventional computer languages.  Thus, we will need new methods to program them.

To develop such methods, we introduced a simple model called a randomly
assembled computer (RAC), where nano-electronic components are placed
and connected at random on a chip.  Input, output, and control lines
connect the chip to the outside world.  By manipulating the control
voltages, different functions, $f$, can be implemented.  Programming is achieved
by minimizing an error function $E_f(\vec{c})$, that depended on the values of the
controls $\vec{c}$.  This is a potentially dif{fi}cult optimization
problem since in general $E_f$ may be a very nonlinear function in a high
dimensional space.  Furthermore, since the RAC was treated as a black box,
the functional form for E was unknown.  The only information available
is the input/output pairs that resulted for a given set of controls $\vec{c}$.
It should be emphasized that this model is used only for algorithm development.
In this paper, we are not advocating a particular hardware or physical design.
In fact, these methods are general and should be applicable to circuits with
a wide range of internal structures.

We considered two methods to optimize the error function: conventional
simulated annealing and a more recently developed multi-agent approach.
Multi-agent optimization methods were found to be well-suited for this class
of problems.  These are adaptive methods based on learning algorithms.
Unlike more conventional techniques, such as simulated annealing, these methods
are distributed, use reinforcement learning, and have the possibility of
assigning different objective functions to different optimization processes
or agents.  This collection of features can result in a substantial increase
in convergence, especially for high dimensional problems.

One of the main question we sought to address was whether an arbitrary function
can be programmed on a RAC.  To help answer this question, we began by considering
programming small RACs as logic gates, such as INV and XOR.
Simulation results suggest that a surprisingly wide-variety of functions can be
implemented on a RAC of {\em suf{fi}cient} size.
For logic gates, we found that that size was on the order of $N\geq18$ diodes
to ensure a greater than $90\%$ success rate.
We found simulated annealing adequate for programming RACs of this size.

For larger RACs and more complicated functions, however, multi-agent
methods were found to be essential for successful programming. We
considered two-bit adders and multipliers programmed on RACs with 
(N=100,K=50).  The space of functions with 4 inputs and 3 outputs is very large 
and picking one particular function out of that large number of
possibilities is a nontrivial problem.  Nonetheless, we found a large
fraction of RACs could do just that using multi-agent methods.  Simulated
annealing, on the other hand, was shown to be not nearly as effective.

Finally, we considered issues related to fault tolerance, reprogrammability,
and reliability.  Electronic components at the nanoscale are expected to be
very sensitive to randomness, and indeed, some degree of randomness may be
unavoidable in certain types of components, such as single electron
transistors.  This potential intrinsic randomness could seriously impact
the reliability of the components, and is thus a serious concern.  RACs,
on the other hand, are quite robust with respect to such randomness.  The
same adaptive methods that dealt with the black box nature of a RAC could adapt
equally well to other potential sources of randomness. This could include
intrinsic randomness of components, random assembly, as well as the unexpected
appearance of faults.  Furthermore, once a RAC is programmed, it could be as
easily re-programmed, and reused potentially in a different application.

\bibliographystyle{plain}

\clearpage

\clearpage

\clearpage

\begin{table}[thb]
\centering
\begin{tabular}{|c|c|c|c|} \hline
$I_1$ & $I_2$ & $I_1+I_2$ & $I_1*I_2$ \\  \hline
0 0 & 0 0 & 0 0 0 & 0 0 0 \\
0 0 & 0 1 & 0 0 1 & 0 0 0 \\
0 0 & 1 0 & 0 1 0 & 0 0 0 \\
0 0 & 1 1 & 0 1 1 & 0 0 0 \\
0 1 & 0 0 & 0 0 1 & 0 0 0 \\
0 1 & 0 1 & 0 1 0 & 0 0 1 \\
0 1 & 1 0 & 0 1 1 & 0 1 0 \\
0 1 & 1 1 & 1 0 0 & 0 1 1 \\
1 0 & 0 0 & 0 1 0 & 0 0 0 \\
1 0 & 0 1 & 0 1 1 & 0 1 0 \\
1 0 & 1 0 & 1 0 0 & 1 0 0 \\
1 0 & 1 1 & 1 0 1 & 1 1 0 \\
1 1 & 0 0 & 0 1 1 & 0 0 0 \\
1 1 & 0 1 & 1 0 0 & 0 1 1 \\
1 1 & 1 0 & 1 0 1 & 1 1 0 \\
1 1 & 1 1 & 1 1 0 & 0 0 1 \\ \hline
\end{tabular}
\caption{
\label{tab:truthadd}
Truth table for a 2-Bit Adder ($O=I_1+I_2$)
and a 2-Bit Multiplier ($O=I_1*I_2$)
}
\end{table}

\clearpage

\clearpage

\clearpage

\clearpage

\end{document}